**Effect of heat treatment on the current induced dynamic mixed metal-insulator phase in needle-like VO$_2$ single crystals.**

L. Patlagan, G. M. Reisner and B. Fisher[*] Physics Department, Technion, Haifa 32000, Israel

## Abstract

The Insulator–Metal-Transition adjoined by a structural transition of VO$_2$ is induced above room temperature (340 K) by heating or self-heating. A steep resistance-jump of up to five orders of magnitude occurs at this transition in high quality, unstrained single crystals. Insulating domains sliding in the sense of the electric current within the metallic background were found so far exclusively in the current induced mixed metal-insulator phase of VO$_2$ single crystals; it is known for a long time that their uniformity and speed as function of current density are very sensitive to crystal quality. The high energetical cost of domain emission is the focus of our present investigations. In this Communication we report on the surprising behavior of a needle-like VO$_2$ single crystal. Several I-V closed loops traced at room temperature concurrently with video recording of the crystal under the microscope, were followed by R(T) measurements during three slow heating-cooling cycles between room temperature and above 340 K, followed in their turn by an additional set of I-V measurements and video recordings. The results show that the slow cycling through the transition using external heat had a healing effect on the reproducibility of R(T) while increasing the activation energy of conduction in the insulating state and in reducing the damping term in the domains' sliding velocity. The intriguing result of this set of measurements was that the energy cost of the domain emission, was *higher after healing* than prior to it.

1. Introduction

VO$_2$ is the prototype material for Insulator–Metal -Transition (IMT); the steep resistance jump (up to 5 orders of magnitude in pure, unstrained single crystals) is close to room temperature ($T_{IMT}$ = 340 K). An updated state of the art of the very active and promising field of VO$_2$ research may be obtained from References [1-5] and references therein.

The transition is from a low temperature monoclinic phase to a high temperature rutile phase; the monoclinic structure of lower symmetry removes the degeneracy between the four equivalent directions of the tetragonal structure and may give rise to twin domains. [6,7] Usually, high quality needle-like VO$_2$ single crystals are single domain (twin boundaries cost energy) and their activation energy of conduction is high ($\Delta \approx 0.45$ eV [8]).

This transition may be induced by various types of external heating or by self-heating. The resistance of a pure VO$_2$ crystal under applied d.c. voltage at an ambient temperature far below T$_{IMT}$ decreases due to Joule heating; upon increasing current the voltage reaches a maximum value, V=V$_{max}$, and a current-controlled negative differential resistance – ((CC)NDR) regime sets in. In this regime the I-V characteristic is stable for R$_L$ ≥ |dV/dI|$_{max}$ (R$_L$ - the load resistance) [9] and smooth. With raising current the temperature of the sample continues to increase (and the resistance to decrease) until T$_{IMT}$ is reached. Unlike the case of external heating, which may lead to full metallization, self-heating under either steady or unsteady state conditions does not lead to full metallization, because the low power (P=IV) in the metal cannot cover the losses at T$_{IMT}$. Instead, a mixed metal-insulator state sets in, consisting of static or dynamic domain patterns. [10]

The change in the optical properties at the IMT allows distinguishing between the insulating and the metallic phases in the mixed state of the sample under the microscope. Under proper orientation and in the rare cases when twin domains are present, one type of insulating twins appears as the negative of the other type (dark and bright) due to polarization by reflection, [11] while the color of the rutile phase is dim.

The most impressive effect—so far unique to VO$_2$ single crystals—is the sliding of narrow insulating domains within a metallic background in the positive sense of the electric current.[10,12] The closest scenario to the sliding domains in VO$_2$ single crystals was recently provided by the electric-field induced mixed insulator-metal state of Ca$_2$RuO$_4$; there, the metallic phase expands upon increasing currents by sliding of the metal-insulator interphase in the sense of the current.[13] In VO$_2$ crystals the sliding domains are emitted from unstable insulating static domains and their sliding is driven by exchange of Peltier heat with latent heat at their boundaries. The sliding velocity, u, as function of current density, J, is given by: [10]

$$u = \frac{\Pi(J-J_o)}{L} \quad (1)$$

where $\Pi$ is the Peltier coefficient of the VO$_2$ I-M couple, L- the latent heat per unit volume of metallic VO$_2$ and $\frac{\Pi}{L}J_0$ is a damping term caused by finite heat conduction.

At the onset of the mixed M-I state of a needle–like VO$_2$ single crystal the trace of the power P(=IV) as function of current bends but continues to increase. The frequency, f, of sliding domains' emission increases with P and the energy per emitted domain is given by $\Delta E = dP/df$. This process which emulates the firing of information by a neuron (domain emission) and propagation of information along an axon (sliding domains along the needle-like crystal) is as yet very inefficient at low frequencies. [14] There was some evidence that the efficiency increases with increasing frequency in ranges that cannot be measured reliably. Our present investigations on VO$_2$ single crystals are focused on finding the reason behind the high energy cost of sliding domain emission and in trying to reduce it.

Routinely, the investigation of a new needle-like VO$_2$ single crystal starts with its inspection under the microscope, I-V tracing and video recording at ambient temperature. Crystals having regular behavior in the dynamic mixed metal-insulator regime are chosen for a further test which consists of R(T) measurements up to T$_{IMT}$ and beyond. Those with steep IMT and MIT transitions, reproducible R(T) upon heating and cooling and high activation energy ($\Delta$) are chosen for further investigations. This procedure, applied on a sample labeled D1(23), revealed a series of discrepancies from our quality criteria (see below) but instead of being discarded it was subjected to further investigation. The measurements dictated by the unusual yet puzzling behavior of this crystal, produced the results shown below that hint on a possible more efficient domain emission even at low frequencies.

2. Material and Methods

Single crystals of VO$_2$ were grown by self-flux-evaporation from V$_2$O$_5$ powder at 980°C under flowing nitrogen for a week. D1(23), the crystal investigated in this work, was a shiny needle with rectangular cross-sections, as were many of the crystals from batch D1 used in the past; its dimensions were 0.0042×0.0028×0.12 cm$^3$. I(V) and R(T) were measured in the four-probe-two-contacts configuration using indium-amalgam dots for contacts; these contacts allow the crystals to shake or bend freely over the structural phase transition being held only by surface tension. I-V measurements were carried out at ambient temperature (22°C+ ~6°C due to microscope illumination). I-V measurements, carried out mostly, but not exclusively, under steady state conditions (RL≥|dV/dI|$_{max}$) [9] were recorded on a YEW type 3036 X-Y recorder. The videos, concurrent with I-V tracing, were recorded at fixed currents, each for about 20 s, using the camera of a smartphone and processed using the movie maker software. The emission frequency, f, was determined by the number of domains crossing a fixed point in the sample per unit time; the sliding velocity, u, a much more complicated measurement, was determined from the time of passage of a domain boundary between two fixed points in the sample. Usually, each data point for u represents an average over 5 periods Δt measured around various instances along the video clip. The range of the f measurements depends on a factor ≤ 1 that changes the videos' speed read by the software. The penalty for extending the range of the measurements is a decreasing color-contrast of the image with the decrease of this factor. With the available software we can measure reliably frequencies only up to ~10 s$^{-1}$.

### 3. Results and discussion.

Figure 1(a) shows the I-V loops of crystal D1(23) measured at ambient temperature with four different load resistances (R$_L$). The onset of NDR along all traces is between 63-64 V. The highest three load resistances provide steady state conditions, while for the lowest (250 kΩ), NDR is on the verge of instability (see green traces). The onset of the mixed state, which occurs above the onset of NDR, is marked by the appearance of sliding domains. The open circles mark all the points where tracing of I-V was stopped for video recordings on the increasing or decreasing current. The duration of each recording was around 20 seconds. Figure 1(b) shows images snipped at various instances from the video clip recorded for I=0.3 mA (see green circle on (a)). In a few cases the dynamic mixed state consisted by bright followed by dark sliding domains, an event observed in twinned crystals [7], but for the first time in a crystal from batch D1. The upper two images and the lowest two, were snipped 0.5 seconds apart close to the start and in the middle of the video-clip, respectively; these images show bright domains followed by dark ones sliding from left to right (which corresponds to the positive sense of the electric current), within the metallic background along the needle- like crystal. They also show that the sliding velocity is constant.

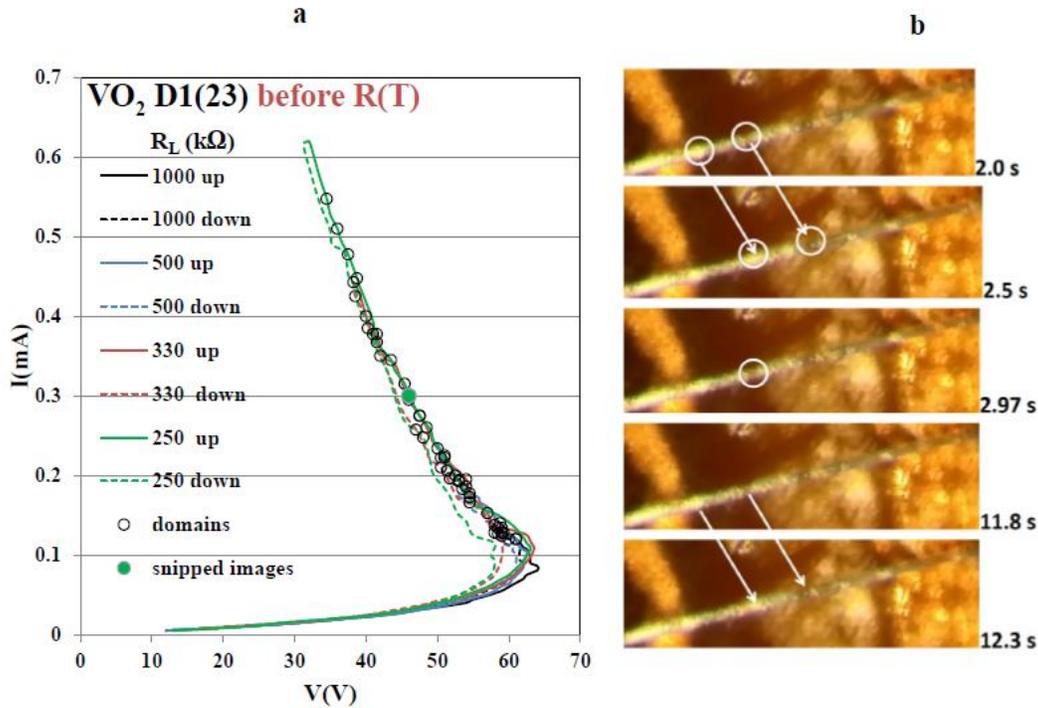

Figure 1. (a) I-V characteristics of crystal D1(23) for $R_L$= 1000, 500, 330 and 250 k$\Omega$; (b) images snipped at various instances from the video-clip recorded for I=0.3 mA (see video-clip 1). Pair of sliding twin domains are encircled for t=2.0 and 2.5 s.

The I-V measurements were followed by slow R(T) measurements. Figure 2 (a) shows semi-log plots of the resistivity ($\rho$) versus 1000/T obtained from these measurements, using the dimensions of the crystal (0.0042×0.0028×0.12 cm$^3$). The upper trace shows relatively large values of $\rho$ and its slope corresponds to a relatively low activation energy of conduction, $\Delta 1$. Following an erratic behavior of the measured voltage at IMT, the measurements were stopped and the crystal was cooled to RT. Two more cycles of R(T) measurements, were carried out on this sample upon heating from RT to 355 K and cooling back to RT. Perfect reproducibility of R(T) was attained over the third heating-cooling cycle for which the values of $\rho$ were lower and the activation energy of conduction ($\Delta 2$) was maximal, indicating some kind of healing. The same data of $\rho$ versus 1000/T appear in Figure 2(b) on linear-linear scale, emphasizing the irreproducibility of the second heating-cooling cycle in contrast to the highly reproducible third cycle. It should be stresses that whatever the nature of the imperfection leading to the heating-cooling irreproducibility of traces 1 and 2, its effect on the metallic regime is hidden within not more than one order of magnitude, as indicated by the identical values of the resistivity measured in that regime for traces 2 and 3 following steep jumps over four orders of magnitude.

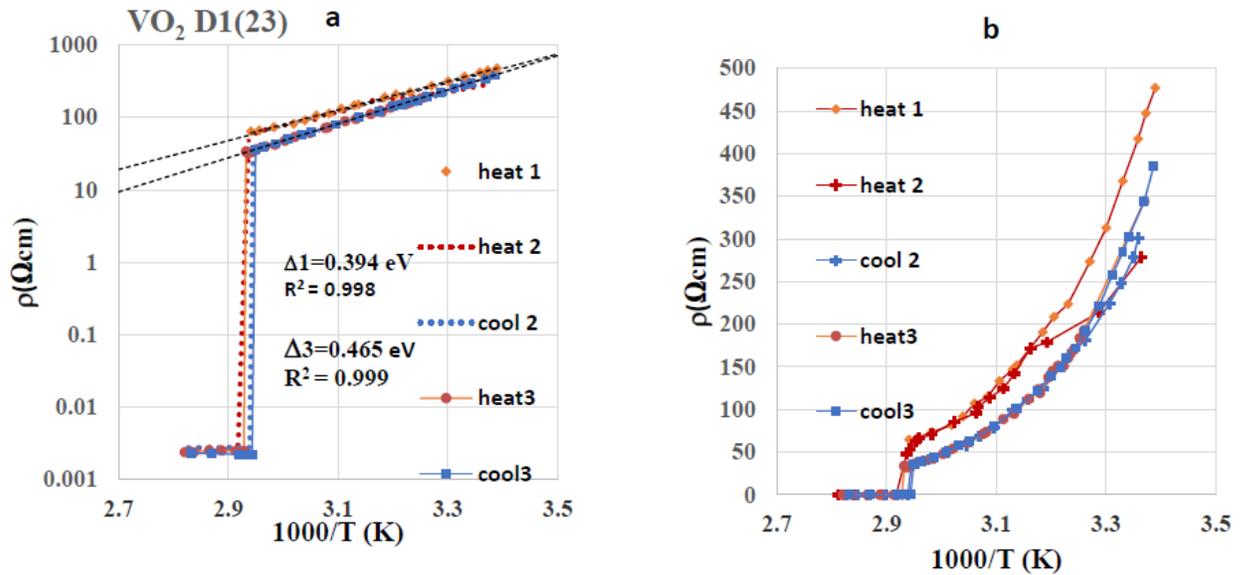

Figure 2. (a) Semi log plot of ρ versus 1000/T of crystal D1(23) over the first heating and over the second and third heating-cooling cycles. (b) same data as in (a) plotted on linear-linear scale which emphasizes the high reproducibility of the third heating-cooling cycle of R(T) in contrast to the irreproducibility of the previous cycles

I-V characteristic measurements concurrent with video recording were carried out on the "healed crystal". Figure 3(a) shows the I-V characteristics for only two different load resistances. It resembles Figure 1(a) but the onset of NDR occurs at a slightly larger voltage than before ($V_{max}$=72-72.6 V). The open circles mark all the points where tracing of I-V was stopped for video recordings upon increasing or decreasing current. As expected, the sliding domains appear above the onset of NDR. Figure 3(b) shows images snipped at various instances from the video clip recorded for I=0.315 mA. This Figure is typical for narrow insulating domains (of one type) sliding at constant velocity along the metallic background of the needle in the positive sense of the electrical current.

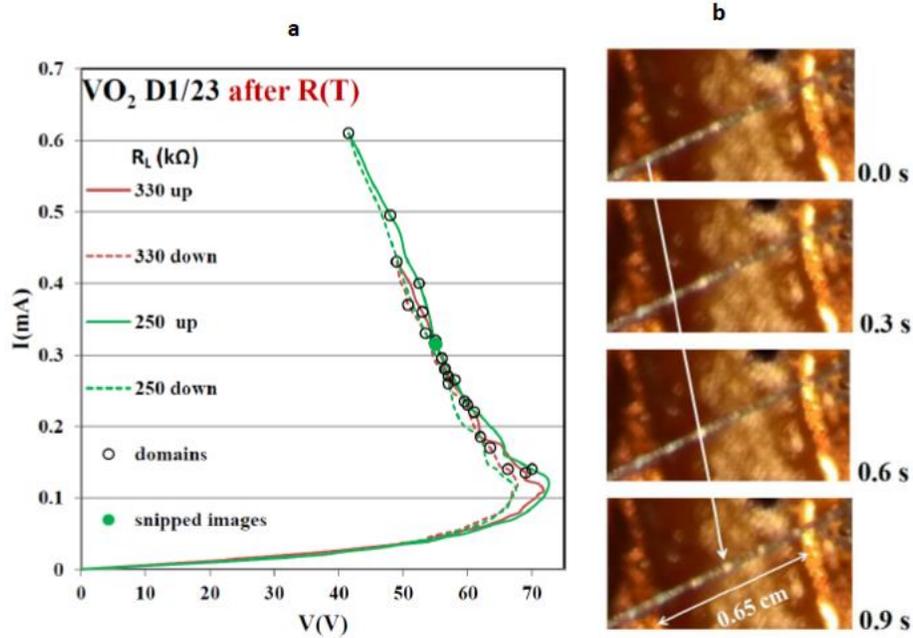

Figure 3. (a) I-V characteristics of crystal D1(23) for $R_L$= 330 and 250 k$\Omega$; (b) images snipped at various instances from the video-clip recorded for I=0.315 mA (see video-clip 2).

The frequency of domains' emission -f and their sliding velocity – u, were measured from the two sets of video-clips recorded (before and after healing) at fixed currents. For the sake of comparison, the two sets of results are superimposed on the same graphs shown below using gray symbols for data obtained "before" and black symbols for data obtained "after" healing.

The measured sliding velocities –u versus current densities J are shown in Figure 4(a) (axis at left). Straight lines fit fairly well the experimental data both before and after "healing". The corresponding slopes of the fitted lines are 0.0018 cm$^3$/(sA) (before) and 0.0016 cm$^3$/(sA) (after). The maximal value of the Peltier coefficient of VO$_2$ measured in the past (0.37 V) [10] and the latent heat per unit volume of VO$_2$ calculated using Reference [15] (240 J/cm$^3$) lead to a ratio of

Π/L=0.00154 cm³/(sA) which is close but somewhat lower than both results. Interestingly, the straight line fitted to the black circles extrapolates to the origin; showing that healing removed completely the damping term in u(J). While the spatial periodicity of the domains, u/f, (axis at right) seems to decrease with increasing J before "healing" it tends to reach a constant value with increasing current for the "healed" crystal. In Figure 4(b), which deals with the energetics of the process, a significant difference between "before" (gray symbols) and "after healing" (black symbols) is shown. Within the same range of currents, (see Figures 1(a) and 3(a)) while P (axis at left) rises linearly to a good approximation with increasing f for the healed crystal (black dots), it decreases with increasing f for the sample before R(T). The traces of $\Delta E(f)$ at the bottom of Figure 4(b) (axis at right) were obtained from the derivatives of the fitted lines to the data of P(f), before and after healing. Accordingly, the energy per emitted domain $\Delta E=dP/df$ (axis at right) is a *constant* for the healed sample but *decreases* with increasing frequency for the sample before healing. For $f=4s^{-1}$, $\Delta E(before)=1mJ$ and $\Delta E(after)=4mJ$. $\Delta E(before)$ continues to decrease to 0.4mJ for $f=9s^{-1}$. The order of magnitude of these values are consistent with those of $\Delta E$ reported in Table 1 of Reference [14].

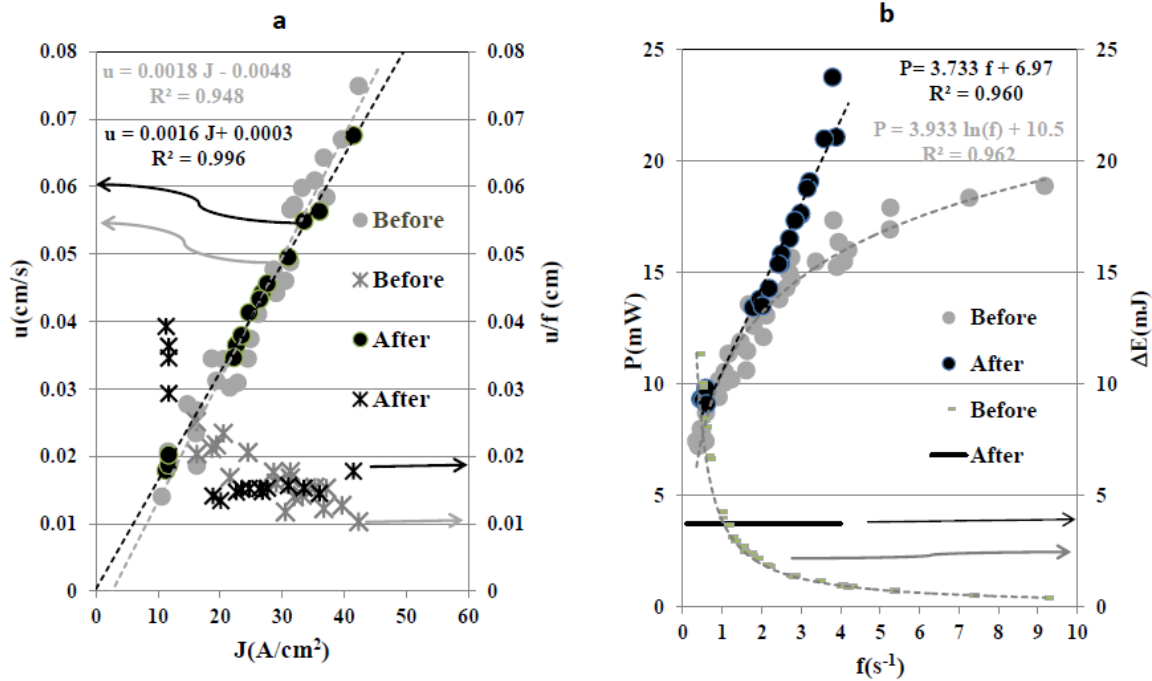

Figure 4. (a) Sliding velocity of domains -u versus current density –J, before (gray symbols) and after (black symbols) "healing" (axis at left); fitted formulas from trend-lines, at the top of Figure. Corresponding spatial periodicities - u/f versus current density (axis at right). (b) Power (P=IV) versus emission frequency –f before (gray symbols) and after "healing" (black symbols); fitted formulas at top of Figure. Emission energy- ΔE versus f calculated from the fitted formulas for P(f) (axis at right).

## 4. Summary and conclusions

In addition to the outstanding IMT transition of $VO_2$ samples of various morphologies, high quality $VO_2$ single crystals poses a unique property: their current induced dynamic mixed metal-insulator regime consists of narrow insulating domains sliding along the metallic background in the positive sense of the electric current. While the sliding process is fairly clear, the emission process is far from being understood. Experimental results, obtained from video recording concurrent with I-V tracing show that in the low frequency regime that can be measured and analyzed using fairly simple equipment and software, the energy cost per domain emission is very high. The crystal D1(23) provided some surprising results that hint toward improving the efficiency of domain emission in the low frequency regime. The crystal exhibited fairly normal behavior over several I-V loops and video recordings except for eventual emission of twin domains. The following slow R(T) measurement, from RT up to IMT and beyond, were very irreproducible indicating the presence of imperfections. Reproducibility was reached over the third heating-cooling cycle indicating healing; the healing consisted also in significant resistivity reduction at all temperatures and increase of the activation energy of conduction. Repeated I-V tracing and video recording that followed the R(T) measurements showed only minor improvements of the sliding process (decrease of the damping term of the sliding velocity and absence of emitted twin domains). The great surprise was that before healing the energy per emitted domain (ΔE=dP/df) decreased significantly with increasing frequency, while after heat treatment it remained large and constant (see Figure 4b). This finding might stimulate further investigation of $VO_2$ crystals with controlled imperfections or doping, either in the bulk [12,16] or at its emitting edge.

**References.**